\documentclass[11pt]{article}
 \pdfoutput=1
\usepackage{jheppub}

\usepackage{amsmath}
\usepackage{verbatim}   
\usepackage{subfigure}  
\usepackage{acronym}

\usepackage{amsfonts}
\usepackage{amssymb}
\usepackage{mathrsfs}
\usepackage{graphicx}
\usepackage{multirow}
 \usepackage{slashed}
 \usepackage{epsfig,multicol,bbm}
 \usepackage{url}

\newcommand{\newc}{\newcommand}
\newc{\gsim}{\lower.7ex\hbox{$\;\stackrel{\textstyle>}{\sim}\;$}}
\newc{\lsim}{\lower.7ex\hbox{$\;\stackrel{\textstyle<}{\sim}\;$}}
\newc{\gev}{\,{\rm GeV}}
\newc{\mev}{\,{\rm MeV}}
\newc{\ev}{\,{\rm eV}}
\newc{\kev}{\,{\rm keV}}
\newc{\tev}{\,{\rm TeV}}
\def\ln{\mathop{\rm ln}}

\newc{\mz}{M_Z}
\newc{\mpl}{M_*}
\newc{\mw}{m_{\rm weak}}
\newc{\nr}[1]{N^c_R{}_{#1}}
\usepackage{amsmath}

\def\beq{\begin{equation}}
\def\eeq{\end{equation}}
\def\bea{\begin{eqnarray}}
\def\eea{\end{eqnarray}}
\def\bitem{\begin{itemize}}
\def\eitem{\end{itemize}}
\newcommand{\bec}{\begin{center}}
\newcommand{\eec}{\end{center}}
\newcommand{\kahler}{K\"{a}hler }

 \newcommand{\mb}{\boldsymbol}
 \newcommand{\der}{\mathrm{d}}

  \newcommand{\GeV}{{\mathrm {GeV}}}

\def\bar#1{\overline{#1}}

\def\inv{^{\raise.15ex\hbox{${\scriptscriptstyle -}$}\kern-.05em 1}}
\def\lbar{{\lower.35ex\hbox{$\mathchar'26$}\mkern-10mu\lambda}} %lambda bar

\let\<=\langle
\let\>=\rangle

\let\+=\uparrow

\let\la=\lambda

\def\OO{\mathcal{O}}

\begin{document}

%%%%%%%%%%%%%%%%%%%%%%%%%%%%%%%%%%%%%%%%%%%%%%%%%%%%%%%%%%%

\title{Precision Unification in $\lambda$SUSY with a 125~GeV Higgs}

\author[a]{Edward Hardy,}
\emailAdd{e.hardy12@physics.ox.ac.uk}
\author[a,b]{John March-Russell,}
\emailAdd{jmr@thphys.ox.ac.uk}
\author[a,c]{James Unwin}
\emailAdd{unwin@maths.ox.ac.uk}
\affiliation[a]{Rudolf Peierls Centre for Theoretical Physics,
University of Oxford,\\
1 Keble Road, Oxford,
OX1 3NP, UK}
\affiliation[b]{Stanford Institute for Theoretical Physics, Department of Physics,\\
 Stanford University, Stanford, CA 94305, USA}
\affiliation[c]{Mathematical Institute,
University of Oxford,
24-29 St Giles, \\
Oxford,
OX1 3LB, UK}

\date{\today}

\abstract{
It is challenging to explain the tentative 125 GeV Higgs signal in the Minimal Supersymmetric Standard Model (MSSM) without introducing excessive fine-tuning, and this motivates the study of non-minimal implementations of low energy supersymmetry (SUSY). A term $\lambda\mb{SH_uH_d}$ involving a Standard Model (SM) singlet state $\mb{S}$ leads to an additional source for the quartic interaction raising the mass of the lightest SM-like Higgs. However, in order to achieve $m_h\approx125~\rm{GeV}$ with light stops and small stop mixing, it is necessary for $\lambda\gtrsim0.7$ and consequently $\lambda$ may become non-perturbative before the unification scale. Moreover, as argued by Barbieri, Hall, {\em et al.}~low fine-tuning prefers the region $\lambda\sim1-2$, leading to new or non-perturbative physics involving $\mb{S}$ below the GUT scale (`$\lambda$SUSY' models). This raises the concern that precision gauge coupling unification, the prime piece of indirect experimental evidence for low energy SUSY, may be upset. Using the NSVZ exact $\beta$-function along with well motivated assumptions on the strong coupling dynamics we show that this is not necessarily the case, but rather there exist classes of UV completions where the strong-coupling effects can naturally correct for the present $\sim3\%$ discrepancy in the two-loop MSSM unification prediction for $\alpha_s$. Moreover, we argue that in certain scenarios a period of strong coupling can also be beneficial for $t-b$ unification, while maintaining the small to moderate values of $\tan\beta$ preferred by the Higgs mass.}

\hfill \vspace{-5mm} OUTP-12-10P

\maketitle

%%%%%%%%%%%%%%%%%%%%%%%%%%%%%
%%%%%%%%%%%%%%%%%%%%%%%%%%%%%

\section{Introduction}

Low energy supersymmetry (SUSY) provides a compelling solution to the hierarchy problem and the observation of the large top mass and improved gauge unification, combined with strong constraints on alternative frameworks from electroweak precision measurements \cite{pdg}, have, since the 1990s, made SUSY the leading candidate to supplant the Standard Model (SM). However, the tentative Higgs signal around $125~\GeV$ reported by ATLAS and CMS \cite{LHC}  is problematic for the simplest implementation: the Minimally Supersymmetric Standard Model (MSSM). Difficulties arise since in the MSSM the Higgs quartic coupling is determined by the electroweak gauge couplings, which results in a tree-level mass of the lightest Higgs state less than the mass of the $Z$ boson.

Common approaches to raising the mass of the lightest Higgs state are through large loop corrections, new contributions to the quartic Higgs coupling, or via level repulsion due to mixing between the Higgs and a SM singlet state. Probably the most studied possibility involves stop squarks significantly heavier than the top quark leading to significant contributions from the stop loops. However as the stop mass increases this reintroduces fine-tuning of the electroweak scale in tension with the requirement that SUSY solves the hierarchy problem. Specifically, it is difficult to obtain $m_h\approx125~\GeV$ with natural stop masses ($m_{\widetilde{t}}\lesssim500~\GeV$) in the MSSM unless there is near-maximal mixing between $\widetilde{t}_L$ and $\widetilde{t}_R$ \cite{MSSM}, which in turn requires very large $A$-terms that are difficult to generate in models naturally solving the SUSY flavour problem. 

In this letter we shall focus on the well motivated approach of introducing a new source for the quartic Higgs interaction via the superpotential term $\lambda \mb{SH_uH_d}$, which involves a new SM singlet state $\mb{S}$, as found the Next-to-Minimal Supersymmetric Standard Model (NMSSM). Including this as well as leading loop corrections leads to contributions to the mass of the lightest SM-like Higgs state of the form\footnote{For simplicity, and motivated by minimal constructions, we shall assume that $\widetilde{t}_1$ and $\widetilde{t}_2$ are approximately degenerate; our conclusions are not substantially altered upon relaxation of this assumption. Throughout we shall consider only models in which $A$-term contributions are negligible.}
\begin{equation}
\label{EQ1}
m_h^2\simeq m_Z^2\cos^2 2\beta+\frac{3m_t^4}{4\pi^2v^2}\log\left(\frac{m_{\widetilde{t}}^2}{m_t^2}\right)
+\lambda^2v^2\sin^2 2\beta~.
\end{equation}
For sizeable $\lambda\gtrsim0.6$ the new NMSSM contribution provides the dominant correction to the Higgs mass and one can obtain $m_h\approx125~\rm{GeV}$ whilst maintaining natural stop masses and small stop mixing. Moreover, the NMSSM is far from an {\it{ad hoc}} solution, since in addition to providing a possible mechanism for raising the Higgs mass, the principle motivation of the framework is to provide a solution to the well-known $\mu$-problem of the MSSM \cite{Prime}.
It is notable that if the coupling $\lambda\gtrsim0.7$ at the weak scale then it will run non-perturbative before the unification scale. It is then natural to be concerned that such large values may result in undesirable side-effects on precision gauge coupling unification. The aim of this work is to quantify the impact on unification of $\lambda$ running through a period of strong coupling.

Experience with the running of the QED coupling through the QCD strong coupling regime is indicative that non-perturbative dynamics in some sector of a theory is not necessarily disastrous for the evolution of an independent gauge coupling, despite na\"ive expectations based upon cursory examination of the RGEs. Furthermore, arguments based on holomorphy \cite{Novikov:1983uc,Shifman:1986zi,Arkani-Hamed:1997ut, Chang:2004db} lead us to believe that the strong coupling in $\lambda$SUSY should not damage gauge unification.  In this paper we demonstrate that provided the coupling $\lambda$ remains non-perturbative for roughly less than an order of magnitude in energy then this in fact can likely increase the precision of gauge coupling unification, correcting the present $3\%$ discrepancy in MSSM gauge unification  \cite{pdg} due to the strong coupling constant running too fast.\footnote{There exist alternative suggestions to improved the precision of gauge coupling unification e.g. \cite{prec, H&R}.} While it is entirely possible that this present deviation between the predicted $\alpha_s(m_Z)$ and the measured value may be resolved by threshold corrections near the weak or GUT scale \cite{thresh}, there are well motivated cases where these are naturally small \cite{H&R}. We thus find it intriguing that $\lambda$SUSY models may not disturb, but even improve, unification.

This paper is ordered as follows: in Section \ref{S2}  we study how the Higgs mass depends on the parameters $\tan\beta$, $m_{\widetilde{t}}$ and $\lambda_0$ (the weak scale value of $\lambda$) and determine the values of these which result in a lightest SM-like Higgs boson at $m_h\approx125~\GeV$. Further we identify the parameter regions which result in $\lambda$ running non-perturbative before the unification scale and discuss how the scale of strong coupling depends on these parameters. In Sections \ref{S3} and \ref{S4} we demonstrate that running through a region in which $\lambda$ becomes non-perturbative can improve the precision of unification. We also consider a possible link with the observed hierarchy in up-like to down-like quark masses, especially, $m_t/m_b$. In the concluding remarks we summarise our results and comment on related issues.

%%%%%%%%%%%%%%%%%%%%%%%%%%%%%%%%%%%
%%%%%%%%%%%%%%%%%%%%%%%%%%%%%%%%%%%

\section{The 125 GeV Higgs in the NMSSM and \texorpdfstring{$\lambda$}{Lg}SUSY}
\label{S2}

To solve the $\mu$-problem of the MSSM the superpotential term $\mu\mb{H_u}\mb{H_d}$ is replaced\footnote{
In $\lambda$SUSY an explicit $\hat{\mu}\mb{H_u}\mb{H_d}$ term is often added, which is taken to be a small PQ-breaking term.}  in the NMSSM by a trilinear interaction $\lambda\mb{SH_uH_d}$ involving a dynamical SM singlet chiral superfield, $\mb{S}$, and the $\mu$-term is reintroduced upon $\mb{S}$ acquiring a vacuum expectation value (VEV). Possible mechanisms for generating a VEV for $\mb{S}$ in the context of $\lambda$SUSY are discussed in \cite{Barbieri:2007tu}. The introduction of  $\mb{S}$  leads to possible new terms in the superpotential 
\begin{equation}
\mathcal{W}=\mathcal{W}_{\rm{MSSM}} +\lambda\mb{S}\mb{H_u}\mb{H_d}+\xi\mb{S}+\mu^{\prime}\mb{S}^2+\kappa\mb{S}^3~.
\end{equation}
Note that some additional symmetries must be imposed in order to remove the dangerous tadpole term $\xi\mb{S}$ (unless the field $\mb{S}$ is composite with suitably low compositeness scale) and in simplified scenarios it is often assumed that the cubic term $\kappa \mb{S}^3$  is also forbidden. Note that if the trilinear term is allowed in the superpotential the RGEs imply that $\kappa$ quickly evolves to small values at lower energies \cite{Barbieri:2007tu} and thus we shall neglect the cubic term henceforth.\footnote{Sizeable $\kappa$ at the weak scale would result in $\lambda$ running faster, becoming non-perturbative at a lower scale.}

\begin{figure}[t!]
\begin{center}
\includegraphics[height=47mm,width=70mm]{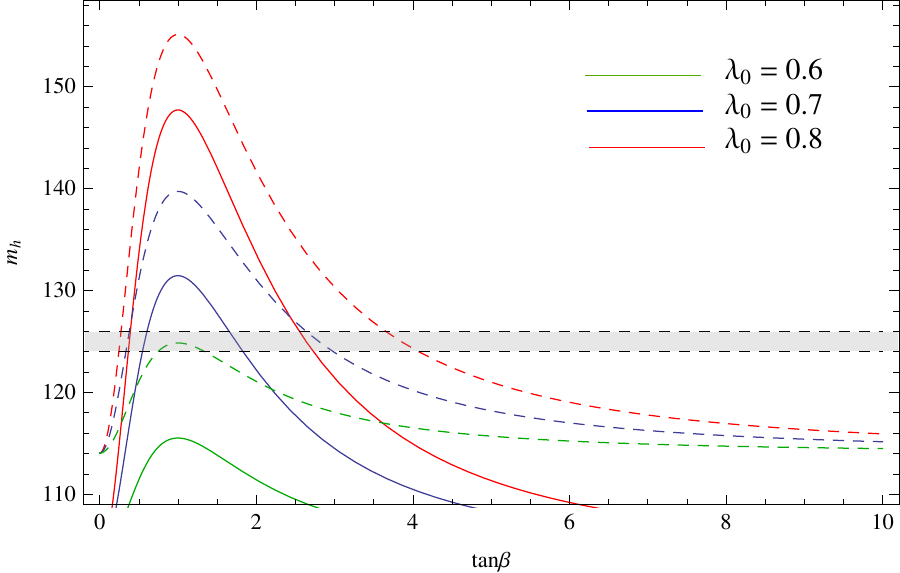}
\includegraphics[height=47mm,width=70mm]{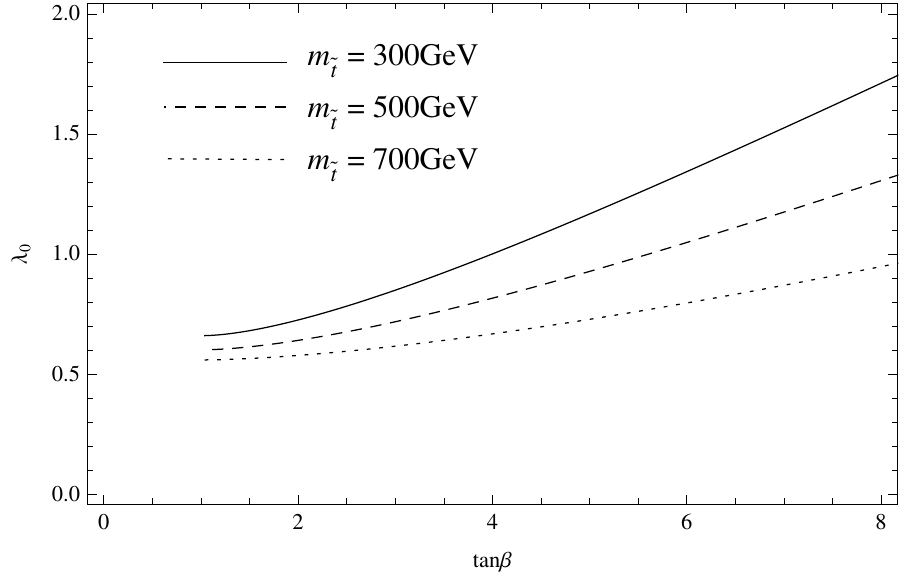}
\caption{{\bf Left.} The variation of the SM-like Higgs mass as a function of $\tan\beta$ for $m_{\widetilde{t}}=300~\rm{GeV}$ (solid curves) and $m_{\widetilde{t}}=500~\rm{GeV}$ (dashed curves) and different values of the starting (weak scale) coupling $\lambda_0$ as indicated. The shaded region corresponds to the possible Higgs signal at 124-126 GeV.  {\bf Right.} The relationship between $\tan\beta$ and $\lambda_0$ which gives $m_h=125~\GeV$ for different stop masses. For $m_{\widetilde{t}}\lesssim500~\rm{GeV}$ this requires $\lambda_0\gtrsim0.65$ and $\lambda$ may run non-perturbative before $M_{\rm{GUT}}$.}
\label{Fig1}
\end{center}
\end{figure}

%%%%%%%%%%%%%%%%%%%%%%%%%%%%%%%%%%%

The leading corrections to the tree-level Higgs mass come from the $F$-term associated with $\lambda\mb{S}\mb{H_u}\mb{H_d}$ and the stop loops, as given in eq.~(\ref{EQ1}). Thus the physical mass of the lightest SM-like Higgs scalar depends on $\tan\beta$, the couplings $\lambda$ and $\kappa$, and the stop mass $m_{\widetilde{t}}$. To give an idea of the dependence we use eq.~(\ref{EQ1}) to calculate the mass of the lightest SM-like Higgs, following \cite{Hall:2011aa}, as a function of $\tan\beta$ for differing values of $m_{\tilde{t}}$ and $\lambda_0$, defined as the value of the coupling $\lambda$ at the weak scale,\footnote{We have neglected two-loop contributions which generally increase the Higgs mass by a few GeV.} this is shown in the left panel of Fig.~\ref{Fig1} (see also \cite{Cao:2012fz}). Observe that $m_h=125$ GeV can not be obtained for $\lambda_0=0.6$ in the case that $m_{\tilde{t}}\lesssim500~\rm{GeV}$. 

At low $\tan\beta$ increasing $\lambda_0$ in the NMSSM leads to a reduction in the amount of fine-tuning and allows for smaller stop masses \cite{Barbieri:2006bg, Hall:2011aa}. In $\la$SUSY models mixing between the singlet and the Higgs is used to lower $m_h$, due to level repulsion  \cite{Hall:2011aa, Jeong:2012ma}, allowing a larger value of $\lambda_0\sim2$ whilst obtaining the desired Higgs mass and consequently leading to a significant reduction in the fine-tuning. (Experimental constraints on models with large $\lambda_0$ have been discussed in \cite{Cao:2008un}.) Alternatively, if the Higgs-singlet mixing is small then $m_h\approx125$ GeV can be obtained with natural stop masses and without stop mixing for somewhat smaller values of $\lambda_0$. However, with light stops and small mixing one requires $\lambda_0\gtrsim0.7$ and the coupling will generally run non-perturbative before the GUT scale.\footnote{
Following Hall {\em et al.} \cite{Hall:2011aa},~we conservatively neglect singlet-Higgs mixing which would reduce the mass of the lightest SM-like Higgs. As we are concerned here with the scenario in which the coupling $\lambda$ is large and the stops are light, higher order corrections to the Higgs mass involving stop loops are small. We consider only models in which $A$-term contributions are negligible, corrections to the Higgs mass due to moderate stop mixing $\delta_X$ compared to the correction $\delta_\lambda$ due to $\lambda\mb{SH_uH_d}$ is $\frac{\delta_X}{\delta_\lambda}\sim\frac{0.068}{\lambda^2\sin^2 2\beta}$ (taking $X_t\simeq m_{\tilde{t}}$).
In models of interest to us here  $\lambda\gtrsim0.7$ and $\tan\beta$ is small ($\sim2$), giving
 $\delta_X/\delta_\lambda\lesssim0.2$ and for larger values of $\lambda$ ($\sim2$) the $\delta_X$ correction is further suppressed.}

In Fig.~\ref{Fig1} ({\em{left}}) the curves with $\lambda_0=0.7,0.8$ have two values of $\tan\beta$ which satisfy $m_h=125~\rm{GeV}$, the lower solution, however, requires $\tan\beta<1$ and such low values are theoretically disfavoured as they result in the top Yukawa running non-perturbative before the unification scale - in the NMSSM $\tan\beta\gtrsim1.5$ is required in order to preserve perturbative {\it SM couplings} up to the unification scale (by adding additional matter in $\mb{5+\bar{5}}$ pairs one can allow  $\tan\beta\gtrsim1$  \cite{Barbieri:2007tu, Masip:1998jc}).\footnote{Although models in which non-SM-singlet states
such as the Higgs doublets or $\bar u_3$ are composite states are of interest (see e.g. \cite{Agashe:2005vg}, in the non-SUSY case), in this work we consider the simplest case in which only SM-singlet states are composite and have large interactions at some scale.}   Consequently, there is a definite relation between $\lambda_0$ and $\tan\beta$ depending only on $m_{\widetilde{t}}$ which we display in Fig.~\ref{Fig1} ({{\em right}}).  We observe that a Higgs in the signal region can be obtained for a range of parameters, with, in many cases, $\lambda$ becoming strongly coupled before the unification scale.

%%%%%%%%%%%%%%%%%%%%%%%%%%%%%%%%%%%

\begin{figure}[t!]
\begin{center}
\includegraphics[height=47mm,width=70mm]{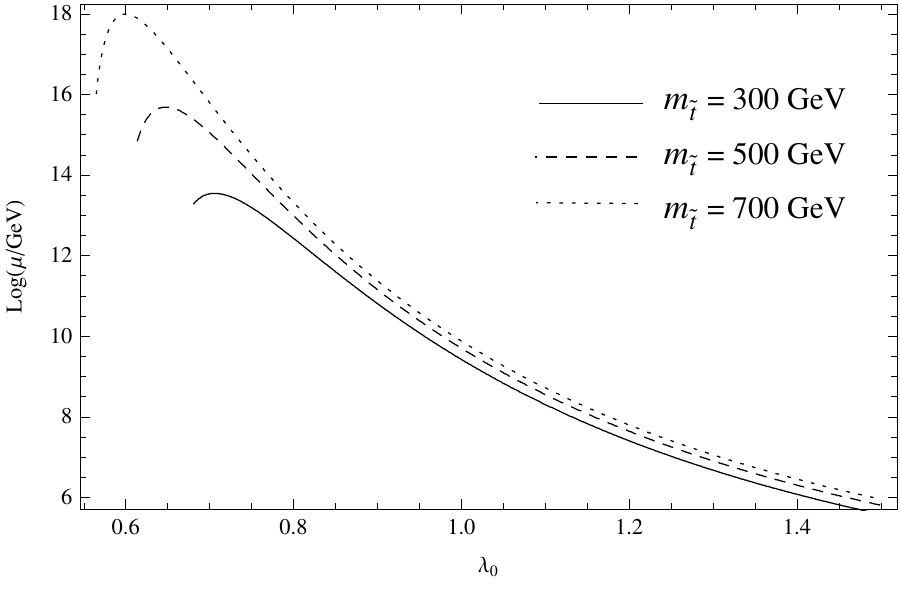}
\includegraphics[height=47mm,width=70mm]{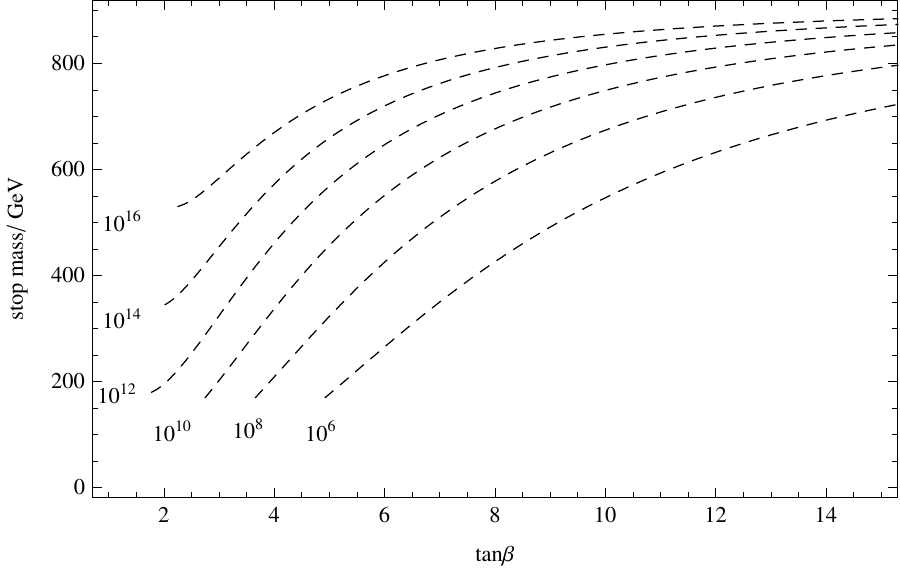}
\caption{{\bf Left.} The strong coupling scale for $\lambda$ against $\lambda_0$ for various stop masses. Note (with our assumptions) for $m_{\widetilde{t}}\lesssim500~\GeV$ the coupling $\lambda$ runs non-perturbative before $M_{\rm{GUT}}$. We fix $\tan\beta$ such that $m_{h}=125~\GeV$ and $\tan\beta>1.5$. {\bf Right.} Contour plot showing the dependence on $\tan\beta$ and $m_{\widetilde{t}}$ of the strong coupling scale for $\lambda$, displaying contours for scales $\geq10^6$ GeV only. We fix  $\lambda_0$, the weak scale value of $\lambda$, such that $m_{h}=125~\GeV$.}
\label{Fig2}
\end{center}
\end{figure}

%%%%%%%%%%%%%%%%%%%%%%%%%%%%%%%%%%%

In Fig.~\ref{Fig2} we use the one-loop RGE evolution of $\lambda$  (see e.g.~\cite{Ellwanger:2009dp}) to study the parameter dependence of the scale $\mu$ at which $\lambda$ becomes strongly coupled,  which we define as $\lambda(\mu)\sim\sqrt{4\pi}$ (the results are insensitive to the exact definition). Judicious parameter choices, with the inclusion of some mixing, can result in perturbativity of $\lambda$ up to the unification scale for models with $m_{\widetilde{t}}\lesssim500 ~\GeV$. With small mixing, it can be seen from Fig.~\ref{Fig2} that for $m_{\widetilde{t}}\lesssim500~\GeV$ (with our previously stated assumptions), the coupling $\lambda$ always runs non-perturbative before the unification scale. Depending on the parameter choices this can occur anywhere from $10^5~\GeV$ to just below the unification scale. As noted previously, large $\lambda_0$ can reduce the fine-tuning, hence $\la$SUSY provides a well motivated scenario in which we expect either new physics to appear before the non-perturbative scale, or the theory to run through a strong coupling regime.

%%%%%%%%%%%%%%%%%%%%%%%%%%%%%
%%%%%%%%%%%%%%%%%%%%%%%%%%%%%

\section{Running through strong coupling}
\label{S3}

If $\lambda$ runs to strong coupling then there are two conceivable scenarios. The theory may remain in a quasi-conformal strong coupling regime all the way to the GUT scale (which need only be an order of magnitude higher in energy scale in some cases). Alternatively, after a brief period of strong coupling the degrees of freedom may recombine such that the theory reverts back to a weakly coupled system with the IR fields composites of the UV degrees of freedom. Examples of the first case occur in Randall-Sundrum-like models where the IR brane scale is the strong coupling scale, while explicit realisations of the second scenario can arise, for example, in \cite{Nakayama:2011iv} and the Fat Higgs models \cite{fat, Chang:2004db}.
In both cases the period of strong coupling will modify gauge coupling unification. As we shall see, however, it will not necessarily destroy successful unification and in some cases can enhance the precision. From the perspective of unification we are most interested in the case where the SM gauge coupling $\beta$-function coefficients below and above the strong coupling regime are such as that the ratios of differences $\frac{b_2-b_3}{b_1-b_2}$ are unaltered, thus maintaining the success of SUSY unification at the leading one-loop log-resummed level. An example of this case occurs when the singlet field $S$ is composite but the Higgs fields are fundamental; such a model was constructed in \cite{Chang:2004db}. We will argue, self-consistently, that even though $\lambda$ becomes non-perturbative and $S$ is replaced by some more elementary degrees of freedom, SM gauge couplings remain perturbatively small throughout the strong coupling region and the effect of this regime is of the form of a threshold correction whose sign and size are reliably estimated with not unreasonable assumptions. 

To quantify the effect of the strong coupling period on gauge unification, consider a theory where  $\lambda$ becomes strongly coupled at a scale $\mu_-$ and remains so until some higher scale $\mu_+$ at which the theory UV completes to a more fundamental weakly-coupled theory. The scenario in which the theory remains strongly coupled up to the GUT scale is simply a special case for which $\mu_+$ is identified with $M_{\rm{GUT}}$. Recalling that the holomorphic `Wilsonian' gauge kinetic function is renormalised only at one-loop, the strongly coupled sector modifies the MSSM $\beta$-functions solely through the anomalous rescaling of matter fields needed to canonically normalise the \kahler potential. The effect on the running is encapsulated in the NSVZ $\beta$-function for the
gauge-coupling evolution in a supersymmetric Yang-Mills-matter theory \cite{Novikov:1983uc,Shifman:1986zi}:
\begin{equation}
\beta_{g_a}\equiv\frac{\der g_a}{\der t}=\frac{g_a^3}{16\pi^2}b_a~,
\label{3.1}
\end{equation}
with $t=\ln\left(Q/M_{\rm{GUT}}\right)$ and
\begin{equation}
b_a=-\frac{3C_2(G_a)-\sum_R T_a(R)\left[1-\gamma_R\right]}{1-\frac{g_a^2}{8\pi^2}C_2(G_a)}~,
\label{2}
\end{equation}
where the index $R$ labels all matter representations, $T_a(R)$ is the quadratic index of $R$, $C_2(G_a)$ is the quadratic Casimir of the group $G_a$ (normalised so that $C_2({\rm{SU}}(N))=N$ and $T_2(\square)=\frac{1}{2}$), and $\gamma_R$ are the matter field anomalous dimensions.  The use of the supersymmetric $\beta$-function is justified as the non-perturbative scales we consider
are much larger than the scale of soft supersymmetry breaking $\sim$TeV.  In eq.~(\ref{3.1}) $g_a$ is the canonically normalised `physical' gauge coupling of the 1PI effective action, and not the holomorphic coupling, a change which leads to the non-trivial denominator (see \cite{Arkani-Hamed:1997ut} for details). In the cases of interest the factor $\frac{g_a^2}{8\pi^2}C_2(G_a)$ is small as the SM gauge couplings
$g_a$ will remain perturbative, hence the denominator may be approximated by 1 if we work to one-loop order in SM gauge couplings in the mixed gauge coupling-$\gamma_R$ terms (but non-perturbative in $\lambda$). 

Outside of the strong coupling region the anomalous dimensions, $\gamma_R$, are loop suppressed and small for all fields, and the one-loop $\beta$-functions are those of the MSSM
\begin{equation}
b_{a}^{(0)}\simeq-\left(3C_2(G_a)-\sum_R T_a(R)\right)~,
\end{equation}
while in the region of strong coupling $b_a$ picks up a new contribution due to non-SM-singlet fields with large anomalous dimensions
\begin{equation}
\Delta b_a^{\rm{(SC)}}\simeq-\sum_R T_a(R)\gamma_R~.
\label{4}
\end{equation}
In the NMSSM the only fields with SM gauge charges that are coupled directly to the strongly interacting sector are $H_u$ and $H_d$, and therefore these fields alone pick-up large anomalous dimensions at the point that the coupling $\lambda$ becomes large. However, the large anomalous dimensions for the Higgs fields will feed into the Yukawa interactions and, as a result, the top Yukawa may subsequently also develop a large anomalous dimension depending on the size of the strong coupling region and the magnitude of $\gamma_{H_u}$; we shall discuss this in detail shortly.

We make the reasonable assumption that during the period of strong coupling, $\mu_-<\mu<\mu_+$, the anomalous dimensions of $H_u$ and $H_d$ are not $\gg1$ (this assumption will be quantified shortly). 
 Hence, calling $g_a$ the `unperturbed' RGE gauge coupling trajectory, i.e.~neglecting corrections due to $\Delta b_a^{\rm{(SC)}}$, the RGEs for the gauge couplings can be approximated as
\begin{equation}
\beta_{g_a}=\frac{g_a^3}{16\pi^2}(b_a^{(0)}+\Delta b_a^{(0)}+\Delta b_a^{\rm{(SC)}})~,
\end{equation}
where $g_a=g_a^{(0)}+\Delta g_a$  is the modified coupling trajectory and the effects of MSSM two-loop diagrams, corrections due to Yukawa interactions and scheme conversion effects are included as an additional perturbation $\Delta b_a^{(0)}$ (which from numerical studies is known to be small in practice, and which we later include). Writing the formal solution to eq.~(\ref{3.1}) as an integral from the IR weak scale to the UV GUT scale we get
\begin{equation}
\int^{g_a(m_Z)}_{g} \frac{ \der g_a}{g_a^{3}}=
\int^{t_{\mathrm{IR}}}_0\frac{b_a\der t}{16\pi^2}~,
\label{5}
\end{equation}
where $g$ is the (normalised) unified coupling at the GUT scale and $t_{\mathrm{IR}}=\left(m_Z/M_\mathrm{GUT} \right)$. The two-loop MSSM and scheme conversion corrections, $\Delta b_a^{(0)}$, are small and therefore induce small finite corrections $\Delta_a^{(0)}$ to the final value of the gauge couplings at the UV scale.  The corrections $\Delta_a^{(0)}$ are independent of $\gamma_R$ to leading order, and thus can
be well-approximated by constant numerical shifts derived from numerical solution of the usual two-loop MSSM RGEs.  As the behaviour of $b_a$ is different in the region of strong coupling, the integration should be partitioned thus
\begin{equation}
\begin{aligned}
\int^{t_{\mathrm{IR}}}_0\frac{b_a\der t}{16\pi^2}= 
&\int^{\ln\left(\frac{\mu+}{M_{\mathrm{GUT}}}\right)}_0\frac{ b^{(0)}_a\der t}{16\pi^2}
+\int_{\ln\left(\frac{\mu+}{M_{\mathrm{GUT}}}\right)}^{\ln\left(\frac{\mu-}{M_{\mathrm{GUT}}}\right)}\frac{ (b_a^{(0)}+\Delta b_a)\der t}{16\pi^2}
+\int_{\ln\left(\frac{\mu-}{M_{\mathrm{GUT}}}\right)}^{\ln\left(\frac{m_Z}{M_{\mathrm{GUT}}}\right)}\frac{ b^{(0)}_a\der t}{16\pi^2}+\frac{1}{2}\Delta_a^{(0)}~.
\end{aligned}
\end{equation}

To parameterise the effects of the strong coupling, we approximate $\gamma_R$ by a constant over the entire region $\mu_-<\mu<\mu_+$ and their usual perturbative
value everywhere else. This, of course, is not meant to be a realistic description of the behaviour of $\gamma_R$ in the strong coupling regime.  Nevertheless, in a self consistent perturbative
expansion in the SM gauge couplings, the leading effect of the large anomalous dimensions is expressible purely as an integral of $\sum_R T_a(R)\gamma_R$ over the strong coupling regime, the
sign and size of which we can parameterise in terms of a constant over $\mu_-<\mu<\mu_+$.  Specifically, from eq.~(\ref{5}) we then obtain
\begin{equation}
\frac{16\pi^2}{g_a^2(m_Z)}
=\frac{16\pi^2}{g^2} + \left[L_a+\Delta^{\mathrm{SC}}_a+\Delta_a^{\left(0\right)}\right]~,
\label{6}
\end{equation}
\begin{equation}
L_a=b^{(0)}_a\ln\left(\frac{M^2_{\mathrm{GUT}}}{m^2_Z}\right)~,
\end{equation}
and we have used eq.~(\ref{4}) in defining
\begin{equation}
\Delta^{\mathrm{SC}}_a\equiv
-\sum_R T_a(R)\gamma_R\ln\left(\frac{\mu_+^2}{\mu_-^2}\right)~.
\label{Del}
\end{equation}

Only the Higgs sector is directly sensitive to the coupling $\la$, thus we expect only $\Delta_{1,2}^{(\rm{SC})}\neq0$ and $\Delta_{3}^{(\rm{SC})}=0$, up to small corrections. 
The sign of the corrections $\Delta_{1,2}^{\rm{SC}}$ is important to us. In the perturbative $\lambda$ regime the Higgs anomalous dimensions are given at one-loop by
\begin{equation}
\begin{aligned}
\gamma(H_u) &=\frac{1}{32\pi^2}\left(2\lambda^2+6h_t^2-g_1^2-3g_2^2\right)~,\\
\gamma(H_d) &=\frac{1}{32\pi^2}\left(2\lambda^2+6h_b^2+2h_\tau^2-g_1^2-3g_2^2\right)~,
\label{per}
\end{aligned}
\end{equation}
where $h_i$, for $i=t,b,\tau$, are the SM Yukawa couplings. Then from definition eq.~(\ref{Del}) and since $T_{1,2}(H_u,H_d)>0$, both $\Delta^{\mathrm{SC}}_1(H_u,H_d)\leq0$ and $\Delta^{\mathrm{SC}}_2(H_u,H_d)\leq0$. Outside of the perturbative regime we cannot make a rigorous statement as the usual unitarity constraint on the wavefunction renormalisation coefficient, $0\leq Z\leq1$, implies only that (the $\lambda$-dependent pieces of) $\gamma(H_u,H_d)\geq0$ in perturbation theory. Nevertheless, a reasonable expectation, in the cases of most interest to us, where the theory doesn't UV complete to a quasi-superconformal model, is that $\Delta^{\mathrm{SC}}_{1}=\Delta^{\mathrm{SC}}_{2}\leq0$ remains true.
If the theory remains strongly coupled for roughly an order of magnitude, the typical size of the deviation due to strong coupling is $\Delta^{\mathrm{SC}}_a\sim-5$, which is parametrically smaller than the standard size RGE-resummed loop corrections $L_2\approx66$ and $L_1\approx198$.
This allows us to perform expansions in the small quantities $\Delta_a^{(\rm{SC)}}/L_a$ to solve for the modified gauge coupling RG trajectories.

The Higgs anomalous dimensions $\gamma_{H_u}$ and $\gamma_{H_d}$ feed directly into the RGE evolution of the top and bottom Yukawas, respectively, which in the strongly coupled region, to leading order, evolve according to
\begin{equation}
\frac{{\rm d} h_t}{{\rm d}t}~\simeq~\gamma_{H_u}h_t~,
\quad {\rm and}~ \quad
\frac{{\rm d} h_b}{{\rm d}t}~\simeq~\gamma_{H_d}h_b~.
\end{equation}
So far our results have only depended upon the sum of the Higgs anomalous dimensions $(\gamma_{H_u}+\gamma_{H_d})$, since $T_a(H_u)=T_a(H_d)$. Whilst an extrapolation of eq.~(\ref{per}), which gives the perturbative forms of $\gamma_{H_u}$ and $\gamma_{H_d}$, would suggest that $\gamma_{H_u}\simeq\gamma_{H_d}$ for large $\lambda$, in the non-perturbative regime these expressions are no longer reliable and this need not necessarily be the case. From a top-down perspective it is natural that no two operators of the strongly interacting theory not appearing in a single irreducible multiplet of the symmetry group of the UV theory should have the same operator dimension, thus implying that $\gamma_{H_u}\neq\gamma_{H_d}$ in general. In fact any dynamical explanation of the MSSM flavour structure must violate a na\"ive extrapolation of the perturbative expression so that the anomalous dimension of the bottom quark mass term (and first two generation fermion mass terms) is large while that of the top remains small, for example as discussed in \cite{Nelson:2000sn}.

If $h_t$ is not to become non-perturbatively large itself (likely implying that $\bar u_3$ and/or $Q_3$ are also composite states), we require that  $\gamma_{H_u}<\gamma_{H_d}$, with $\gamma_{H_u}$ bounded above by 
\begin{equation}
\gamma_{H_u}\lesssim\frac{0.5}{\ln\left(\frac{\mu_+}{\mu_-}\right)/\ln(10)}~.
\end{equation}
 The difference $(\gamma_{H_u}-\gamma_{H_d})$ allows an interesting possibility, providing an explanation for the hierarchy between up-like and down-like quark masses which does not rely on large $\tan\beta$, as is usually assumed, but instead is due to the greater running of $h_b$ compared to $h_t$, starting from a common value $h_b\simeq h_t\simeq\OO(1)$ at the GUT scale. Specifically, if 
\begin{equation}
(\gamma_{H_u}-\gamma_{H_d}){\rm ln}\left(\frac{\mu_-}{\mu_+}\right)\sim 4~,
\end{equation}
then the observed small ratio $m_b/m_t$ is obtained without resort to $\tan\beta\gg1$. In fact if the Higgs contribution due to $\lambda\mb{SH_uH_d}$ is to raise the Higgs mass to 125 GeV, then $\tan\beta\lesssim10$ is required, as illustrated in Fig.~\ref{Fig1}, so an independent explanation of the top to bottom mass hierarchy is necessary.

Alternatively, if $\gamma_{H_u}\gtrsim0.5$, then the top will also generally develop a sizeable anomalous dimension shortly after the period of strong coupling begins. 
This provides an additional contribution to $\Delta_a^{\rm SC}$:
\begin{equation}
\Delta^{\mathrm{SC}}_a\equiv
-\left(\sum_{R=H_u,H_d} T_a(R)\gamma_R\right)\ln\left(\frac{\mu_+^2}{\mu_-^2}\right)
-\theta(\mu_+-\mu_t)\left(\sum_{R=t,Q} T_a(R)\gamma_R\right)\ln\left(\frac{\mu_+^2}{\mu_t^2}\right)~,
\label{Del2}
\end{equation}
where $\mu_t$ is the scale at which the top Yukawa becomes non-perturbative. Note that in the case that $\gamma_{H_u}\simeq\gamma_{H_d}$ we expect that the top Yukawa runs non-perturbative shortly after $\lambda$, and  therefore $\mu_t\simeq\mu_-$~.
Importantly, since $T_a(t,Q)>0$, and $\gamma_{\overline{u}_3}$ and $\gamma_{Q_3}$ inherit the same sign as $\gamma_{H_u}$ (at least if the leading perturbative results for the sign of $\gamma_{\overline{u}_3}$ and $\gamma_{Q_3}$ hold), these corrections have the same sign as those due to $\gamma_{H_{u,d}}$, and as we shall see shortly, this only results in a slight deflection in the RGE trajectories of the gauge couplings.
%

%%%%%%%%%%%%%%%%%%%%%%%%%%%%%
%%%%%%%%%%%%%%%%%%%%%%%%%%%%%

\section{Effects of strong coupling on SM gauge couplings at \texorpdfstring{$m_Z$}{Lg}}
\label{S4}

Taking the measured low energy gauge parameters $\alpha_{\rm{em}}\big|_{\overline{\rm{MS}}}~,~m_Z\big|_{\overline{\rm{MS}}}~$ and  $\sin^2\theta_w\big|_{\overline{\mathrm{MS}}}~$ as inputs allows a prediction for $\alpha_{S}(m_Z)\big|_{\overline{\rm{MS}}}~$. From eq.~(\ref{6}) these quantities can be expressed as
\begin{align}
\sin^2\theta_W&=
\frac{3}{8}\left[1-\left(b_1^{(0)}-\frac{5}{3}b_2^{(0)}\right)\frac{\alpha_{\mathrm{em}}}{2\pi}\ln\left(\frac{M_{\mathrm{GUT}}}{m_Z}\right)\right]+\Delta^{s_w}~,
\label{8}
\\[3pt]
\alpha_s^{-1}(m_Z)&=
\frac{3}{8\alpha_{\mathrm{em}}}\left[1-\left(b_1^{(0)}+b_2^{(0)}-\frac{8}{3}b_3^{(0)}\right)\frac{\alpha_{\mathrm{em}}}{2\pi}\ln\left(\frac{M_{\mathrm{GUT}}}{m_Z}\right)\right]+\Delta^{\alpha_s}~,
\label{9}
\end{align}
where $\Delta^{s_w}$ and $\Delta^{\alpha_s}$ are corrections to the one-loop form due to two-loop SM corrections, Yukawa interactions, scheme dependent effects and, now, also the effects of running through a regime of strong coupling. 
To study the effect of the period of strong coupling on the SM gauge couplings we write $\Delta_a=\Delta_a^{(0)}+\Delta_a^{\mathrm{SC}}$ where $\Delta_a^{(0)}$ are the standard MSSM values which are known (see e.g.~\cite{Dienes:1996du, Dienes:1995sq})  to be
$(\Delta_{1}^{(0)},\,\Delta_2^{(0)},\,\Delta_3^{(0)})\simeq(11.6,\,13.0,\,7.0)$ and $\Delta_a^{\mathrm{SC}}$ is the additional correction due to running through a period of strong coupling.
The form of the corrections is given by
\begin{equation}
\begin{aligned}
\Delta^{s_w}
&=-\frac{\alpha_{\mathrm{em}}}{4\pi}\left(\frac{1}{1+\frac{5}{3}}\right)\left[\Delta_1-\frac{5}{3}\Delta_2\right],
\\[3pt]
\Delta^{\alpha_s}&=-\frac{1}{4\pi}\left(\frac{1}{1+\frac{5}{3}}\right)\left[\Delta_1+\Delta_2-\left(1+\frac{5}{3}\right)\Delta_3\right].
\label{10}
\end{aligned}
\end{equation}
Expanding the $\Delta_a$ and using the numerical values for the MSSM corrections in order to assess the impact of the corrections due to strong coupling gives
\begin{equation}
\begin{aligned}
\Delta^{s_w}
&\simeq \frac{\alpha_{\mathrm{em}}}{32\pi}\left[5\Delta^{\mathrm{SC}}_2-3\Delta^{\mathrm{SC}}_1+30.2\right],
\\[3pt]
\Delta^{\alpha_s}&\simeq\frac{1}{32\pi}\left[8\Delta_3^{\mathrm{SC}}-3\Delta_1^{\mathrm{SC}}-3\Delta_2^{\mathrm{SC}}-17.8\right].
\label{11}
\end{aligned}
\end{equation}
%

%%%%%%%%%%%%%%%%%%%%%%%

\begin{figure}[t!]
\begin{center}
\includegraphics[height=65mm]{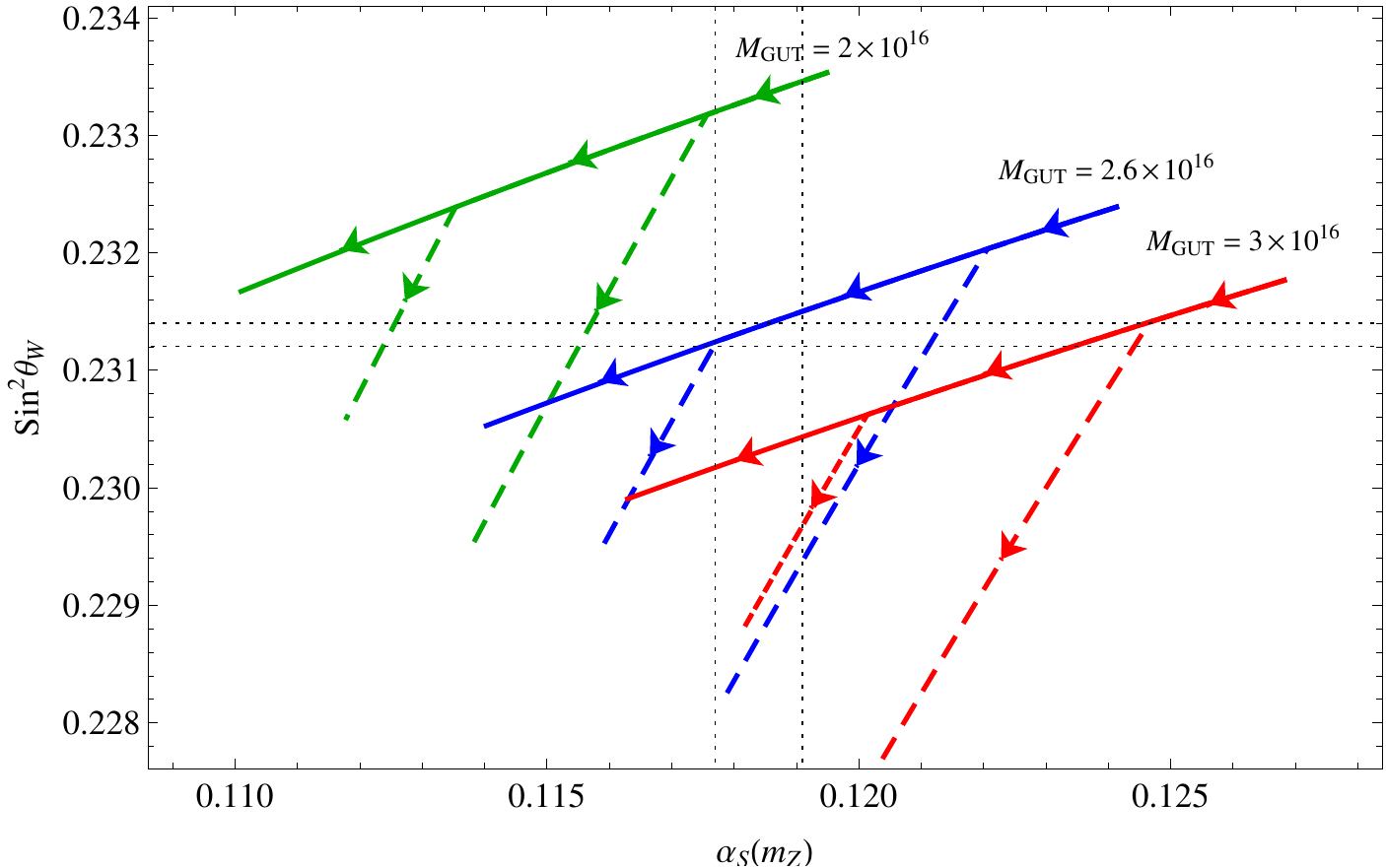}
\caption{The plot shows the effect of $\Delta^{\rm{SC}}_a$  on the predicted values of $\alpha_s(m_Z)$ and $\sin^2\theta_W$ for a range of unification scales $M_{\rm{GUT}}$. The start point of each curve indicates the MSSM value (ie.~$\Delta^{\rm SC}_a=0$) and the arrows indicate the trajectories for increasing values of the quantity $t\equiv(\gamma_{H_u}+\gamma_{H_d}){\rm ln}(\mu_+/\mu_-)$, showing $0\leq t\leq 6$. The default preferred scenario is shown by the solid lines which assume negligible anomalous dimensions for the top states, a self-consistent assumption if $\gamma_{H_u}$ is not too large.  In the case of large $\gamma_{H_u}$ the top Yukawa coupling runs to non-perturbative values leading to large anomalous dimensions for the 3rd-generation states $Q_3$ and ${\bar u_3}$.  Assuming $\gamma_{\overline{u}_3}\simeq \gamma_{Q_3} \neq 0$, the trajectory will be deflected depending on the scale at which $h_t$ becomes non-perturbative, as shown schematically by the heavy dashed lines as this scale is varied over the allowed range. The black dotted lines show the preferred region as indicated by current experimental measurements \cite{pdg} (including errors):  $\sin^2\theta_w\big|_{\overline{\mathrm{MS}}}=0.2313\pm0.001$ and $\alpha_s(m_Z)=0.1184\pm0.007$.}
\label{Fig3}
\end{center}
\end{figure}

%%%%%%%%%%%%%%%%%%%%%%

The low energy gauge parameters are well measured and there is a reasonable level of agreement with the predictions of gauge coupling unification assuming the MSSM spectrum. However, as stated previously there is a $\sim3\%$ deviation between the predictions for $\alpha_s(m_Z)$ from MSSM unification and the measured values \cite{pdg} of  $\sin^2\theta_w\big|_{\overline{\mathrm{MS}}}=0.2313\pm0.001$ and $\alpha_s(m_Z)=0.1184\pm0.007$. In Fig.~\ref{Fig3} we plot the low energy observables as a function of $M_{\rm{GUT}}$ and the quantity $(\gamma_{H_u}+\gamma_{H_d}){\rm ln}(\mu_+/\mu_-)$.  The new corrections entering due to the region of strong coupling have the right sign if, as expected, $\Delta^{\mathrm{SC}}_a\leq0$, and possibly even the correct magnitude, to correct for the discrepancy in MSSM unification.

First we shall consider the scenario in which the anomalous dimension of the top is negligible, as is the case if the anomalous dimension for $\gamma_{H_u}$ is small and it is primarily $\gamma_{H_d}$ which is responsible for deviations in the evolution of the gauge couplings. In this situation it is straightforward to determine the parameter values of a strong coupling regime that gives precision unification; using eq.~(\ref{10}) and eq.~(\ref{8}) yields 
\begin{equation}
\sin^2\theta_W\simeq\frac{3}{8}+\frac{\alpha_{\mathrm{em}}}{16\pi}\left[\frac{1}{2}\left[5\Delta^{\mathrm{SC}}_2-3\Delta^{\mathrm{SC}}_1+30.2\right]-\left(3b_1^{(0)}-5b_2^{(0)}\right)\ln\left(\frac{M_{\mathrm{GUT}}}{m_Z}\right)\right]~.
\end{equation}
Recall, in the (N)MSSM the one-loop $\beta$-function coefficients are $b_1^{(0)}=11$, $b_2^{(0)}=1$ and $b_3^{(0)}=-3$ and that $\alpha_{\rm{em}}=1/127.9$. Substituting $\sin^2\theta_W\simeq0.2313$ leads to
\begin{equation}
\ln\left(\frac{M_{\mathrm{GUT}}}{m_Z}\right)\simeq33.53+\frac{5}{56}\Delta^{\mathrm{SC}}_2-\frac{3}{56}\Delta^{\mathrm{SC}}_1~.
\label{GUT}
\end{equation}
Similarly, from eq.~(\ref{11}) and eq.~(\ref{9}) we obtain 
\begin{small}
\begin{equation*}
\alpha_s^{-1}(m_Z)\simeq
\frac{3}{8\alpha_{\mathrm{em}}}+\frac{1}{16\pi}\left[
\frac{1}{2}\left[8\Delta_3^{\mathrm{SC}}-3\Delta_1^{\mathrm{SC}}-3\Delta_2^{\mathrm{SC}}-17.8\right]
-\left(3b_1^{(0)}+3b_2^{(0)}-8 b_3^{(0)}\right)\ln\left(\frac{M_{\mathrm{GUT}}}{m_Z}\right)\right]
\end{equation*}
\end{small}
and by comparison with eq.~(\ref{GUT}) we have
\begin{equation}
\alpha_s(m_Z)\approx
0.129+5.3\times10^{-3}\times\left[
\frac{3}{7}\Delta^{\mathrm{SC}}_2-\frac{3}{28}\Delta^{\mathrm{SC}}_1-\frac{1}{4}\Delta_3^{\mathrm{SC}}
\right]~.
\end{equation}
Thus in order to obtain the observed value $\alpha_s(m_Z)\approx0.118$ it is required that
\begin{equation}
\Delta_3^{\mathrm{SC}}\approx
8.3-0.43\Delta^{\mathrm{SC}}_1+1.71\Delta^{\mathrm{SC}}_2~.
\label{EQ2}
\end{equation}

In the case that only the Higgses acquire large anomalous dimensions, we have $\Delta^{\mathrm{SC}}_1=\Delta^{\mathrm{SC}}_2$ and $\Delta^{\mathrm{SC}}_3=0$ and hence the GUT scale can be expressed as a function of a single argument
\begin{equation}
M_{\rm{GUT}}\sim 
m_Z\exp\left(33.5-\frac{\Delta^{\mathrm{SC}}_1}{28}\right)~.
\end{equation}
The observed value of $\alpha_s(m_Z)\approx0.118$, given in eq.~(\ref{EQ2}), is obtained for $\Delta_1^{\rm{SC}}=\Delta^{\mathrm{SC}}_2=-6.5$, which corresponds to a unification scale of
$M_{\rm{GUT}}\approx2.6\times10^{16}$  GeV. 
Note that the unification scale is slightly raised compared to the standard MSSM prediction, slightly lengthening the predicted proton lifetime arising from dimension six $X$ and $Y$ gauge boson exchange (see e.g. \cite{rev}), as $\tau_p \propto M_X^4 / \alpha_{\rm GUT}^2$ (in addition, $1/\alpha_{\rm GUT}$ increases slightly in our scenario from $\sim 23.6$ to $\sim 23.8$ for $M_{\rm GUT}=2.6\times10^{16}$, further increasing the proton lifetime, though this is a subdominant effect).
Furthermore, since $T(H_{u,d})\big|_{\rm{U}(1)}=1$ we may write 
\begin{equation}
\Delta^{\mathrm{SC}}_1\sim -2(\gamma_{H_u}+\gamma_{H_d})\ln\left(\frac{\mu_+}{\mu_-}\right)~.
\end{equation}
For example, in the case that $\mu_+/\mu_-\simeq10$ to obtain $\Delta_1^{\rm{SC}}=-6.5$ we require an anomalous dimension of $(\gamma_{H_u}+\gamma_{H_d})\sim1.4$, in accord with our expectation for the effective magnitude of the anomalous dimensions during a regime of strong coupling. If $\mu_+/\mu_-\sim2$ then the required anomalous dimension increases to $(\gamma_{H_u}+\gamma_{H_d})\sim4.6$, still within reasonable values. 

It is likely, however, that if $\gamma_{H_u}{\rm log}\left(\frac{\mu_+}{\mu_-}\right)\gtrsim0.5$ then non-perturbative effects due to the top also affect the evolution of the gauge couplings. The case where these effects turn on quickly is shown as dotted curves in Fig.~\ref{Fig3}. However, for an appropriate choice of $M_{\rm GUT}$ it is clear that  precision unification can be achieved regardless of how quickly the non-perturbative effects due to the top enter, provided the period of strong coupling is not too long. Of course it would be false to claim that a period of strong coupling fixes the discrepancy between the MSSM two-loop prediction of $\alpha_s(m_Z)\sim0.129$ and the measured value, rather, our point is that an epoch of strong coupling (with the theory UV completing in such a way that $\frac{b_3-b_2}{b_2-b_1}$ remains unchanged) is not disastrous for precision unification and may even be advantageous.

Another interesting scenario which realises precision unification via running through strong coupling is the case where the  strong coupling region immediately precedes the GUT scale and $\mu_+$ is identified with this unification scale. In this scenario one need not be concerned if the top Yukawa runs non-perturbative. Such strong coupling unification has been previously argued to have advantages for stabilising the string dilaton and may also have interesting consequences for the SUSY spectrum \cite{Kolda:1996ea}. Note that, in Section \ref{Fig2} we identified the parameter regions in which this situation is realised, for example, from inspection of the right panel of Fig.~\ref{Fig2} we observe that for 500 GeV stops and $\tan\beta\simeq3$, then the strong coupling window starts at $\mu_-\sim10^{15}$ GeV, only an order of magnitude below the GUT scale.

\section{Concluding remarks}
\label{S5}

A Higgs boson as heavy as $125~\GeV$ is difficult to explain in the MSSM, and the NMSSM provides an attractive framework for explaining the tentative Higgs signal. However, as shown in Figs~\ref{Fig1} and \ref{Fig2}, for stop masses light enough that there is not excessive fine-tuning, we require that $\lambda_0\gtrsim0.7$ to obtain the desired Higgs mass and for such large values of $\lambda$ at the weak scale the coupling will generally become strongly coupled before unification. This is even more the case in the $\lambda$SUSY scenario of Refs.~\cite{Barbieri:2006bg, Hall:2011aa} where $\lambda_0 \sim 2$ is argued to substantially reduce low-energy fine-tuning in the electroweak sector.   A coupling becoming strongly coupled before unification raises the concern that successful gauge coupling unification may be adversely affected.  However, on the contrary, we argued in this paper that gauge coupling unification is, in suitable cases, likely improved given a short period of strong coupling.  In these advantageous cases, the strong coupling regime corresponds to a threshold effect of sign and size expected to be of the right order to correct the current 3\% discrepancy between the two-loop MSSM prediction for $\alpha_s(m_Z)$ and its measured value (our final results being given in Section \ref{S4}). 
Moreover, we argued that in scenarios where $\gamma_{H_u}<\gamma_{H_d}$, a period of strong coupling can also be beneficial for $t-b$ unification.

Since the motivation for $\lambda$SUSY is predicated on the 125 GeV Higgs signal, it is worth investigating if other aspects of Higgs phenomenology, particularly the production cross-section and branching ratios, favour the $\lambda$SUSY scenario.
Currently, the branching ratios seem roughly SM-like, however there appears to be an enhancement in the rate $pp\rightarrow H\rightarrow\gamma\gamma$ \cite{Higgs}. It has been argued that the $\gamma\gamma$ signal can be enhanced in the NMSSM and that such an enhancement favours larger values of $\lambda$ \cite{Ellwanger:2011aa}.\footnote{It has also been suggested \cite{Das:2012ys} that the NMSSM can provide an explanation of the tentative Fermi 130 GeV photon line \cite{Fermi}.} As $\lambda$SUSY is a leading mechanism for raising the Higgs mass in a way that reduces fine-tuning, and strong coupling need not adversely affect precision gauge coupling unification, new anomalies arising in the data (such as the tentative signals mentioned above) certainly warrant dedicated studies in the context of $\lambda$SUSY.

\section*{Acknowledgements}

We are grateful to Mads Frandsen, Lawrence Hall, Christopher McCabe, Yasunori Nomura and Kai Schmidt-Hoberg for useful discussions. We would also like to thank the JHEP referee for their useful comments.
JMR and JU would like to thank the Stanford Institute for Theoretical Physics for their hospitality.
EH is supported by an STFC Postgraduate Studentship.
JMR is supported in part by both ERC Advanced Grant BSMOXFORD 228169, and acknowledges support from EU ITN grant UNILHC 237920 (Unification in the LHC era).
JU is grateful for support from the Esson Bequest, Mathematical Institute, Oxford and the Vice-Chancellors Fund, University of Oxford.

\end{document}